\documentclass[sigconf, nonacm,pdfa]{acmart}

\newcommand\vldbdoi{10.14778/3611540.3611636}
\newcommand\vldbpages{4130 - 4137}
\newcommand\vldbvolume{16}
\newcommand\vldbissue{12}
\newcommand\vldbyear{2023}
\newcommand\vldbauthors{\authors}
 
\newcommand\vldbavailabilityurl{}
\newcommand\vldbpagestyle{empty} 

\newcommand{\eg}{\emph{e.g.}}
\newcommand{\ie}{\emph{i.e.}}
\newcommand{\etc}{{\em etc.}}
\newcommand{\eat}[1]{}

\begin{document}
\title{Generations of Knowledge Graphs: \\The Crazy Ideas and the Business Impact}

\author{Xin Luna Dong}
\affiliation{%
  \city{Redmond}
  \state{Washington}
  \postcode{98004}
}
\email{lunadong@gmail.com}

\begin{abstract}
Knowledge Graphs (KGs) have been used to support a wide range of applications, from web search to personal assistant. In this paper, we describe three generations of knowledge graphs: {\em entity-based KGs}, which have been supporting general search and question answering (\eg, at Google and Bing); {\em text-rich KGs}, which have been supporting search and recommendations for products, bio-informatics, etc. (\eg, at Amazon and Alibaba); and the emerging integration of KGs and LLMs, which we call {\em dual neural KGs}. We describe the characteristics of each generation of KGs, the crazy ideas behind the scenes in constructing such KGs, and the techniques developed over time to enable industry impact. In addition, we use KGs as examples to demonstrate a recipe to evolve research ideas from innovations to production practice, and then to the next level of innovations, to advance both science and business.

\end{abstract}

\maketitle

\pagestyle{\vldbpagestyle}
\begingroup\small\noindent\raggedright\textbf{PVLDB Reference Format:}\\
\vldbauthors. Generations of Knowledge Graphs: The Crazy Ideas and the Business Impact. PVLDB, \vldbvolume(\vldbissue): \vldbpages, \vldbyear.\\
\href{https://doi.org/\vldbdoi}{doi:\vldbdoi}
\endgroup
\begingroup
\renewcommand\thefootnote{}\footnote{\noindent
This work is licensed under the Creative Commons BY-NC-ND 4.0 International License. Visit \url{https://creativecommons.org/licenses/by-nc-nd/4.0/} to view a copy of this license. For any use beyond those covered by this license, obtain permission by emailing \href{mailto:info@vldb.org}{info@vldb.org}. Copyright is held by the owner/author(s). Publication rights licensed to the VLDB Endowment. \\
\raggedright Proceedings of the VLDB Endowment, Vol. \vldbvolume, No. \vldbissue\ %
ISSN 2150-8097. \\
\href{https://doi.org/\vldbdoi}{doi:\vldbdoi} \\
}\addtocounter{footnote}{-1}\endgroup

\ifdefempty{\vldbavailabilityurl}{}{
\vspace{.3cm}
\begingroup\small\noindent\raggedright\textbf{PVLDB Artifact Availability:}\\
The source code, data, and/or other artifacts have been made available at \url{\vldbavailabilityurl}.
\endgroup
}

\bigskip
\textit{"Science is to test crazy ideas; engineering is to bring these ideas into business." }
\ \ \ \ \ \ \ \ \ \ \ \ \ \ \ \ \ \ \ \ \ \ \ \ \ \ \ \ \ \ \ \ \ \ \ \ \ \ \ \ \ \ \ \ \ \ \ \ 
{\em -- Andreas Holzinger}

\section{Introduction}
\label{sec:intro}
Since the birth of modern Knowledge Graphs (KGs) around 2007 (in the same year, Yago~\cite{yago}, DBPedia~\cite{dbpedia}, and Freebase~\cite{freebase} were released)~\footnote{There are two knowledge bases before all KGs discussed in this paper, Cyc (\url{cyc.com}) and WordNet (\url{wordnet.princeton.edu}); they are limited in scope and scale because of hand-crafting.}, the area has been broadly researched in a multitude of research communities (to name a few, NLP, IR, Data Mining, Databases, Semantic Web). The industry deployment started about a decade ago, when Google launched {\em Knowledge Panels} in web search in 2012; since then, KGs have been used broadly to support web search (\eg, Google and Bing web search), voice assistants (\eg, Amazon Alexa, Apple Siri, and Google Assistant), and so on, and have made profound business impact.

KGs model the real world in a graph representation, where nodes represent real-world entities or atomic (attribute) values, and edges represent relations between the entities or attributes between entities and atomic values. A piece of knowledge can be considered as a {\em triple} in the form of (subject, predicate, object), such as (Seattle, {\em located\_at}, USA). The {\em data instances} in a KG follow the {\em ontology} as the schema, which in itself is represented in a graph form and can be taken as a part of the KG. The ontology describes entity {\em classes}, often organized in a hierarchical structure and also called {\em taxonomy}, and meaningful relationships between classes. 

KGs can be considered as {\em semi-structured}: on the one hand, it enjoys clean semantics of structured data powered by the rigidity of schemas (\ie, ontology); on the other hand, it embraces the flexibility of unstructured data by allowing easily adding new classes and relationships. An additional advantage of KGs is that it can seamlessly connect a large number of domains through common entities across domains or relationships between domains (\eg, the {\em Movie} and {\em Music} domains can be connected by people who are both actors/actresses and singers, and by the {\em featured\_song} relation). These advantages give KGs a unique position that is both understandable to machines (through ontology) and easy-to-understand by human beings (blessed by the structure), suitable to facilitate understanding in search, question answering (QA), and dialogs, to power recommendation through the graph structure, and to display information for human understanding (in attribute-value pairs), comparison (in tables), and explanation (in paths in the graph).

With the widespread applications of KGs, {\em how to model and capture all valuable knowledge in the world} has emerged as a prominent research area. This paper delves into this subject through the author's journey in the past decade, enriched with extensive scientific research and production deployment experiences gained at esteemed companies like Google, Amazon, and Meta. 

\subsection{Generations of knowledge graphs}
\label{subsec:generations}
\begin{figure}
  \centering
  \vspace{-.5in}
  
  \includegraphics[width=\linewidth]{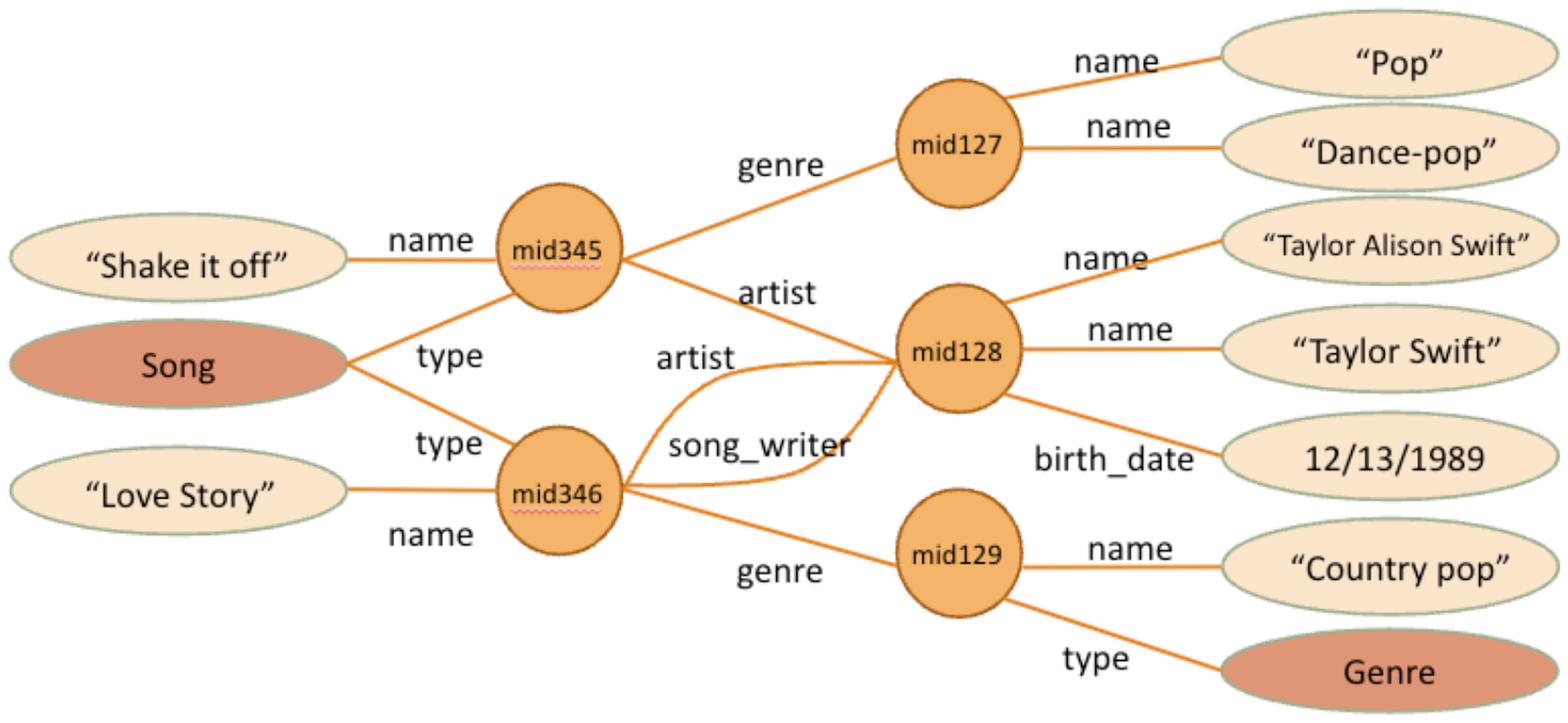}
  
  \vspace{-.4in}

  \raggedright
  (a) An example entity-based KG in the music domain. Nodes are mostly entities, each with an ID.
  
  \vspace{-.6in}
  \centering
  
  \includegraphics[width=0.7\linewidth]{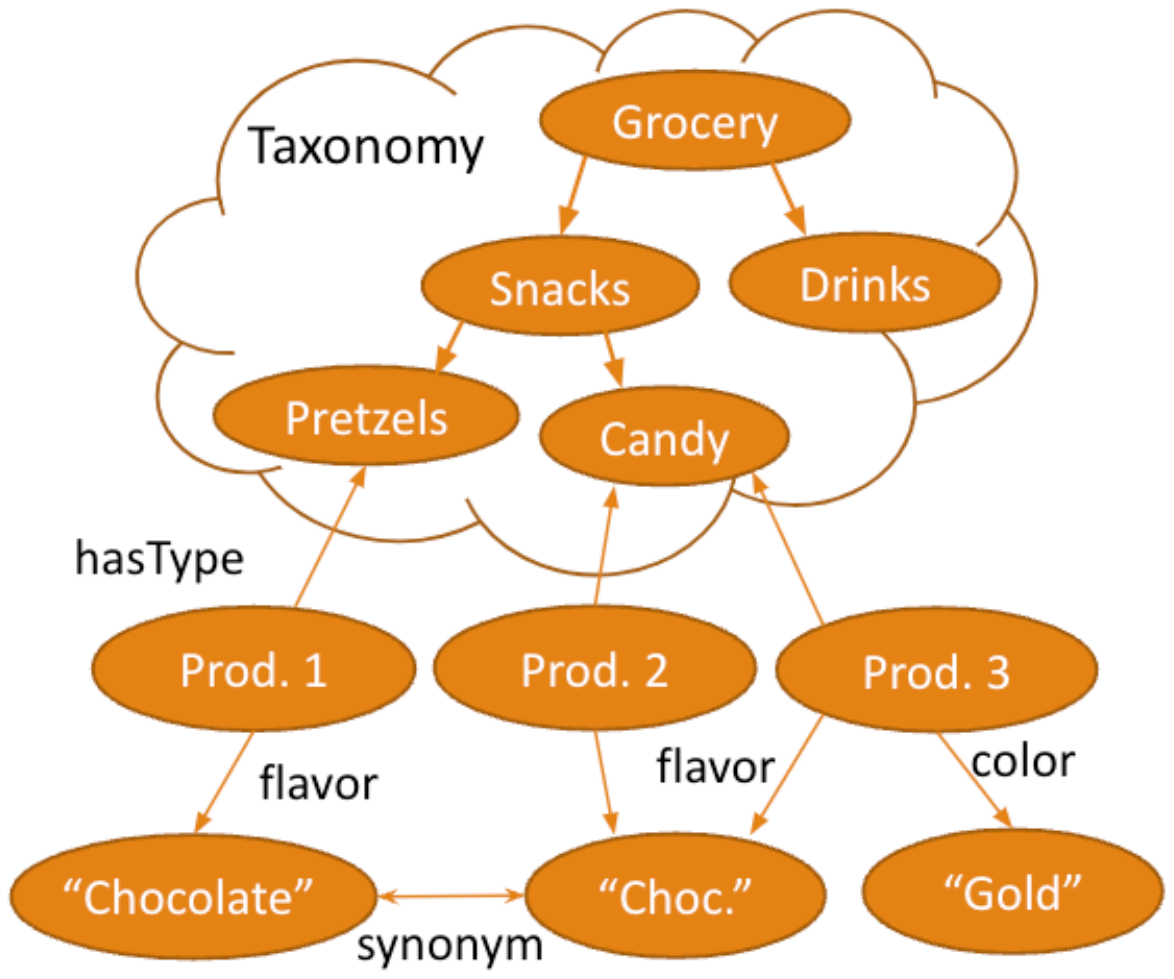}
  
  \raggedright
  \vspace{-.6in}
  
  (b) An example text-rich KG in the product domain. The top depicts the taxonomy, which can be a rich and deep hierarchy. The bottom depicts data instances, where attribute values are mostly texts; as such, it is mostly in the form of a bipartite graph (except edges like "synonym").
  
  \caption{Example knowledge graphs.}
  \label{fig:kg-example}
  \vspace{-.1in}
\end{figure}

In this paper, we discuss a few generations of KGs. The first generation is {\em entity-based} KGs, where both ontology and data are more rigorous, and nodes in the graphs are mostly entities that have one-to-one correspondence with real-world entities (see Figure~\ref{fig:kg-example}(a) as an example). Most well-known generic KGs, such as Yago~\cite{yago} from academia and Google KG~\cite{googlekg} from industry, are entity-based KGs. We discuss this generation in Section~\ref{sec:entityKG}.

The second generation is {\em text-rich} KGs, where ontology and data allow much more ambiguities, and nodes in the graphs are more often just free texts. With the text nodes, the graph is mostly in the form of a bipartite graph, as depicted in Figure~\ref{fig:kg-example}(b). Text-rich KGs are often used to model domains where structure is sparse while ambiguities are abundant, with vague and fluid semantic boundaries between values and even classes, such as {\em Product}, {\em Bio-informatics}, and {\em Health}. Section~\ref{sec:textKG} discusses this generation.

The upcoming generation is not fully shaped yet and we call it {\em dual neural} KGs for now. It encodes knowledge explicitly as triples (as in KGs) and implicitly as embeddings (as in language models). The same piece of knowledge may co-exist in both forms or stay on one side that is more suitable, and there is smooth transition between the two forms to allow harmonic blending. Section~\ref{sec:neuralKG} discusses why we believe {\em co-existing} is the key for success, at least in the near future. 

\subsection{The recipe from innovation to practice}
\label{subsec:cycles}
KG is an area that has witnessed success both in research and in industry. As we discuss the evolution of KGs, we employ it as an example to illustrate the cycle from innovation to production practice, and subsequently to the next round of innovation. This iterative cycle often comprises several stages, each contributing to impacts from initial to profound.

\begin{enumerate}
    \item {\bf Feasibility:} The cycle first starts with a (or a series of) prototype or an experiment, showing the feasibility of a crazy idea, which sometimes seeds a new field. 
    \item {\bf Quality:} The second stage focuses on gradually improving the quality of the solution (a model, an algorithm) to production quality, which enables trustworthy and pleasant user experiences. This is the key stage to land an innovation as a tangible product: unless attaining production quality, a research idea will only remain research.
    \item {\bf Repeatability:} Once we achieve success with the initial product, usually within a limited scope (a few domains, or working under a set of constraint conditions), the next stage is to repeat the success for larger scopes like broader domains. This stage often emphasizes building pipelines to facilitate automation, and employing machine learning (ML) models to minimize manual work. It is a stage leading to much higher business impact.
    \item {\bf Scalability:} Although repeatability could lead to impact enhancement of 1-2 orders of magnitude, it oftentimes falls short of achieving true scalability, demanding impacts of thousands or millions of times. Scalability often necessitates a new set of solutions that substantially reduce costs and eliminate all manual work from the loop.
    \item {\bf Ubiquity:} Finally, ubiquity seeks to maximize the scope of applicability, to encompass long-tail use cases, to remove any underlying assumption in the solutions. Pursuing such solutions often triggers a new round of innovations, initiating the next cycle (sometimes even scalability can lead to the next cycle).
\end{enumerate}

This paper interweaves the discussions of the three generations of KGs and the innovation-to-practice-to-innovation cycle. Through the former, we illustrate how development of techniques leads to larger and larger business impact; through the latter, we shed insights on how the pursuit of large business impact sparks new innovations. Finally, we reflect on critical factors for production success in Section~\ref{sec:reflection}.

\section{Entity-Based Knowledge Graphs}
\label{sec:entityKG}

Entity-based KGs are the most popular KGs in both academia (\eg, Yago~\cite{yago}, DBPedia~\cite{dbpedia}, \etc.) and industry (\eg, Google KG~\cite{googlekg}, Bing Satori KG~\cite{satori}, Alexa KG~\cite{alexakg}, WikiData~\cite{wikidata}). As early as in 2015, it was reported that Google Knowledge Graph shows for about 25\% of all Google search queries~\footnote{\url{https://searchengineland.com/googles-knowledge-graph-may-show-14th-search-queries-212962}}.

There are two characteristics for entity-based KGs. First, the ontology of the KGs is normally manually defined with {\em clear} semantics, where entity types and relationships have few ambiguities or overlaps. For each domain in the ontology, the numbers of entities and relationships are fairly small and thus manageable for manual definition; for example, Freebase contains 52 entity types and 155 relationships in the {\em Movie} domain.

Second, most entities in entity-based KG are {\em named} entities, each corresponding to a real-world entity, such as a person, a university, a movie, a song, and so on. There is rarely overlap between the entities; for example, there are no two persons who are the same, and no two movies that are exactly the same, even if they may share the same name. 

\subsection{{\em Feasibility and Quality:} Knowledge transformation}
\label{subsec:kTransformation}
The seed crazy idea behind entity-based KGs is exactly the idea of modeling the world with entities and relationships. In a sense, that is how human beings understand the world: a child would think about the world as herself, her mom and dad, her friends, her kindergarten, the cartoon she likes, \etc. Now the question becomes, {\em how to identify the entities in the world and discover their relationships from available data sources}? 

\smallskip
\noindent
{\bf Feasibility:} Luckily, the idea of modeling the world with entities and relationships is not new: the DBMS (DataBase Management System) uses {\em ER (Entity Relationship) Diagrams} to visualize the logical structure of the database. Therefore, entities and relationships in KGs can be transformed from structured data such as relational databases. Wikipedia~\cite{wikipedia}, which started in 2001 and describes entities and provides hyperlinks from one entity page to another, conveniently becomes a starting point for collecting knowledge. Wikipedia Infoboxes can be transformed to entities and relationships in a straight-forward way (see an example in \url{https://en.wikipedia.org/wiki/William_Shakespeare}); this spurs successful early KGs such as Yago, DBPedia and Freebase.

\smallskip
\noindent
{\bf Quality:} The high accuracy of Wikipedia data also guarantees reliability of the derived knowledge. So as far as the transformation is carefully curated to ensure semantics correctness, we can achieve high quality. Since 2012, KGs have been used in production as a trustworthy data source, and Wikipedia has been serving as the major source for the majority of generic KGs even now. 

\eat{
\autoref{fig:duck}.

\begin{table*}[t]
  \caption{A double column table.}
  \label{tab:commands}
  \begin{tabular}{ccl}
    \toprule
    A Wide Command Column & A Random Number & Comments\\
    \midrule
    \verb|\tabular| & 100& The content of a table \\
    \verb|\table|  & 300 & For floating tables within a single column\\
    \verb|\table*| & 400 & For wider floating tables that span two columns\\
    \bottomrule
  \end{tabular}
\end{table*}
}

\subsection{{\em Repeatability}: Knowledge integration}
\label{subsec:kIntegration}
With the success of transforming Wikipedia Infoboxes into knowledge, we naturally wish to enrich knowledge from other structured sources, such as {\em IMDb} for movies, {\em MusicBrainz} for music, and {\em Goodreads} for books. These sources may supplement Wikipedia, oftentimes about torso to long-tail entities (in terms of popularity). However, each of these sources organizes its data in a different way, so the next question becomes, {\em how to integrate the knowledge transformed from different structured sources?}

The knowledge integration problem is one form of {\em data integration}, and it needs to resolve three types of heterogeneities~\cite{DS13}:
\begin{itemize}
\item {\bf Schema heterogeneity:} Different data sources may express the same entity type and relationship in different ways (\eg, first name and last name vs. full name). {\em Schema alignment} aligns source schemas with the KG ontology. 
\item {\bf Entity heterogeneity:} Different data sources may represent the same real-world entity with slightly different names, and provide different attribute values (\eg, Xin Dong from Univ. of Washington vs. Xin Luna Dong from Meta). {\em Entity linkage} links such entities such that we have a distinct node in the KG to represent a real-world entity. This problem is even more tricky as different entities may share the same name (thus {\em entity disambiguation}).
\item {\bf Value heterogeneity:} Different data sources may provide different attribute values for the same entity, some of which may be imprecise or out-of-date (refer to the same example for entity heterogeneity). {\em Data fusion} decides among different, and possibly conflicting values, which are correct and up-to-date values. 
\end{itemize}

Among the three problems, schema alignment is mostly done manually to ensure semantics correctness in knowledge transformation; data fusion is less prominent when we restrict knowledge sources to a few authoritative ones. Entity linkage stands out as a critical problem to solve when we link multiple sources, each of which often has millions of entities or more, making manual linkage implausible. 

\begin{figure}
  \centering
  \vspace{-1.2in}
  \includegraphics[width=\linewidth]{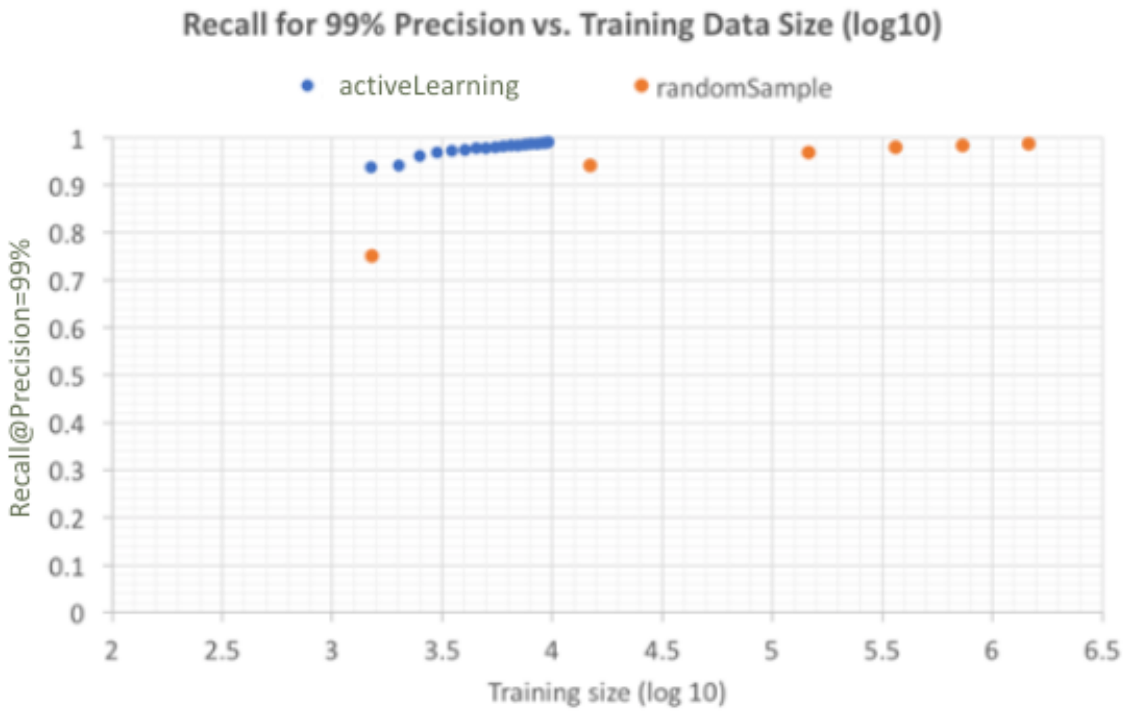}
  \vspace{-1.4in}
  \caption{Entity linkage quality with random forest on movies and people between Freebase and IMDb~\cite{broadgraph}. We are able to achieve over 99\% precision and recall with 1.5M labels. When applying active learning to selectively introduce labels, we can achieve the same quality with 10K labels.}
  \label{fig:linkage}
\end{figure}

Entity linkage is a problem with decades of research, dating back to 1969~\cite{fellegi69}. In practice, tree-based models have been proved to be effective solutions for entity linkage. Figure~\ref{fig:linkage} shows that we can train random forest models that take attribute-wise value similarities as features, and obtain over $99\%$ precision and recall when linking movies and people between Freebase and IMDb. In addition, the figure shows that although very high precision and recall could require a large number of training labels, applying active learning can reduce training labels by orders of magnitude while maintaining similar linkage quality.

Knowledge integration, especially entity linkage, allows us to repeat the success of knowledge collection from Wikipedia to multiple authoritative structured sources. Most of large KGs harvest data from a variety of sources; for example, Freebase takes data from MusicBrainz, NNDB, Fashion Model Directory, \etc.\footnote{\url{https://en.wikipedia.org/wiki/Freebase_(database)}.}

\eat{
\autoref{tab:freq}. 

\begin{table}[hb]
  \caption{Frequency of Special Characters}
  \label{tab:freq}
  \begin{tabular}{ccl}
    \toprule
    Non-English or Math & Frequency & Comments\\
    \midrule
    \O & 1 in 1000& For Swedish names\\
    $\pi$ & 1 in 5 & Common in math\\
    \$ & 4 in 5 & Used in business\\
    $\Psi^2_1$ & 1 in 40\,000 & Unexplained usage\\
  \bottomrule
\end{tabular}
\end{table}
}

\subsection{{\em Scalability}: Knowledge extraction from semi-structured websites}
\label{subsec:kExtraction}
As discussed in Section~\ref{subsec:cycles}, repeatability does not necessarily lead to scalability. We can transform data from tens of structured sources into knowledge and integrate them to create a holistic KG; however, a lot of manual interference is needed and it is hard to scale up to thousands, or even millions of sources. On the web there are numerous {\em semi-structured websites} (\eg, \url{rottentomatoes.com}), where each page represents a {\em topic entity}, and different pages display information in key-value pairs at relatively consistent locations across the pages. These websites are typically populated from large structured data sources, thus serve as good data sources to enrich KGs. If we can automatically extract knowledge from these websites, instead of relying on manual knowledge transformation, we will be able to scale up knowledge collection from structured sources on the web.\footnote{Webtables~\cite{CHW+08} is a special form of semi-structured data.} Three major techniques has been proposed for knowledge extraction from semi-structured data. 

\smallskip
\noindent
    {\bf Wrapper induction:} Wrapper induction, dating back to 1997, takes manual annotations on a few semi-structured webpages from the same website and induces the extraction patterns expressed in XPaths that can apply to the whole website~\cite{KWD97}. This method works because semi-structured websites are normally populated from underlying databases using some templates (\eg, CSS), and wrapper induction reverse engineers the templates. Although wrapper induction can normally obtain high extraction quality (over $95\%$), it still requires annotations on every website so is not {\em truly} web-scale.

\smallskip
\noindent
    {\bf Distantly supervised extraction:} Distantly supervised extraction started in 2014~\cite{DGH+14}; it compares knowledge in existing KGs and data on the semi-structured websites, and generates training data according to the overlaps. Extraction quality is more or less driven by the quality of the training data, so there has been research focused on generating high-quality training data; {\sf Ceres}~\cite{ceres} does so with careful examination of the structure of semi-structured pages and commonality between pages. {\sf OpenCeres}~\cite{open-ceres} further extends this method to annotate (attribute, value) pairs, allowing extracting knowledge for unknown attributes (thus {\em OpenIE}). This class of methods trains a model per website (precisely, for each cluster of webpages in the website that apply the same template), but the whole process is automatic and thus can scale up to a large number of websites. 

   \smallskip
    \noindent
    {\bf GNN-based extraction:} The intuition behind GNN-based extraction is that given a semi-structured webpage, one can fairly easily guess what is the topic entity, and what are the attribute-value pairs, without domain knowledge, and even without necessarily understanding the language (\eg, in foreign language). Systems like {\sf ZeroshotCeres}~\cite{zeroshot-ceres} leverages Graph Neural Network (GNN) to explore both the visual clues and the text semantics, to train one single extraction model for different websites, including even websites in domains where training data do not exist, pushing the boundary of extracting knowledge for {\em unknown unknowns}.

\begin{figure}
  \centering
  \vspace{-.1in}
  \includegraphics[width=0.7\linewidth]{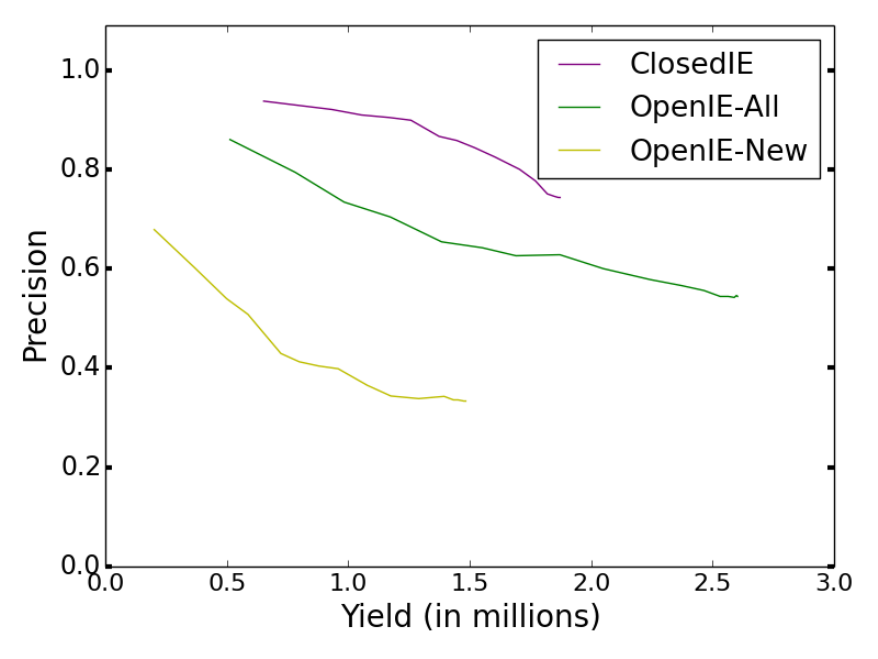}
  \vspace{-.2in}
  \caption{Extraction quality from semi-structured websites~\cite{zeroshot-ceres}, showing that ClosedIE has achieved over 90\% accuracy, whereas OpenIE has shown the promise to increase knowledge, but has much lower accuracy.}
  \label{fig:ceres}
  \vspace{-.3in}
\end{figure}

\smallskip
These three methods are progressively more scalable. As shown in Figure~\ref{fig:ceres}, {\sf Ceres} can achieve over $90\%$ extraction accuracy, thus reaches production quality; whereas extraction for new relations / domains remains in exploratory stages. Finally, knowledge extraction scales up by automating schema alignment, but there will still be needs for entity linkage and data fusion, where there have been plenty of research for large-scale linkage~\cite{JSW+22} and fusion~\cite{DN09, LDL+12b}.

\subsection{{\em Ubiquity:} Web-scale extraction and fusion}
\label{subsec:kFusion}
The web is a huge repository of knowledge and it has been the wish of many researchers to extract knowledge from the whole web to achieve ubiquity of knowledge collection. Well-known projects in this line include {\sf NELL}~\cite{CBK+10} and {\sf Knowledge Vault} (KV)~\cite{DGH+14}. {\sf NELL} focuses on text extraction, whereas {\sf KV} extracts knowledge from four types of web contents: texts, semi-structured data (as discussed in \ref{subsec:kExtraction}), web tables, and HTML annotations (\eg, according to {\em schema.org}). 

To achieve web-scale, we need an efficient way to generate training data to cover various data patterns. Distant supervision~\cite{Brin98, AG00, sofie, MBS+09} is applied for this purpose, but the training data and thus the extractions are often noisy. Various {\em knowledge fusion} techniques are proposed to predict correctness of the extractions, such as PRA (Path ranking algorithm) in NELL~\cite{CBK+10}, deep learning based link prediction in KV~\cite{DGH+14}, and graphical models in KV~\cite{DGE+14}. The graphical models are also used to distinguish extraction errors and source errors, leading to web source trustworthiness evaluation, as in Knowledge-Based Trust~\cite{kbtrust}.

With web-scale knowledge extraction and fusion, NELL extracted 435K knowledge triples and KV extracted ~100M triples with over 90\% confidence (94M from semi-structured websites). It is orders of magnitude smaller than commercial KGs (to compare, at the same time point Freebase contained 637M triples and Google KG contained 18B triples). Although web extraction did not generate a huge volume of knowledge as expected, it led to several important insights. First, entity-based knowledge is mainly structured data, so the best knowledge sources are still structured sources; thus, knowledge transformation (Section~\ref{subsec:kTransformation}) and integration (Section~\ref{subsec:kIntegration}) from well-curated structured sources could be the most effective method to collect high quality knowledge. Second, semi-structured websites are major contributors of high-quality knowledge in web extraction, and they can cover long tail knowledge not covered by major structured sources; this insight inspired further investment on knowledge extraction from semi-structured websites, as described in Section~\ref{subsec:kExtraction}. Finally, we find that texts often embrace knowledge not easily captured cleanly by entities, leading to the next generation of KGs, as we will describe in Section~\ref{sec:textKG}.

\subsection{Summary}
\begin{figure}
  \centering
  \vspace{-.7in}
  
  \includegraphics[width=\linewidth]{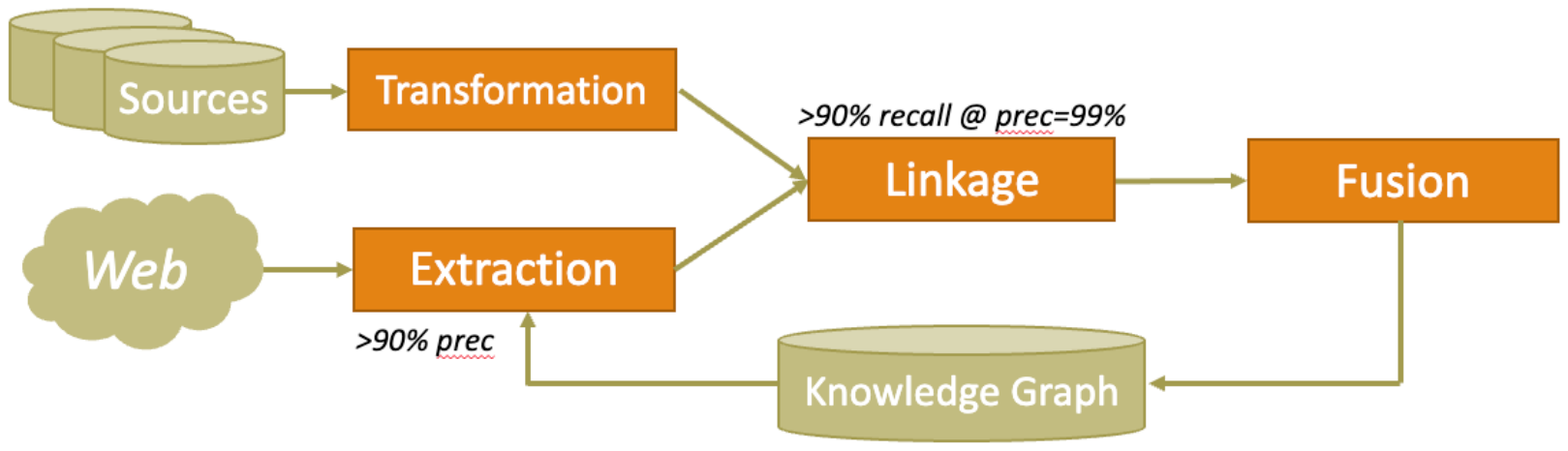}
  
  \vspace{-.75in}
  
  (a)
  
  \vspace{-.75in}
  
  \includegraphics[width=\linewidth]{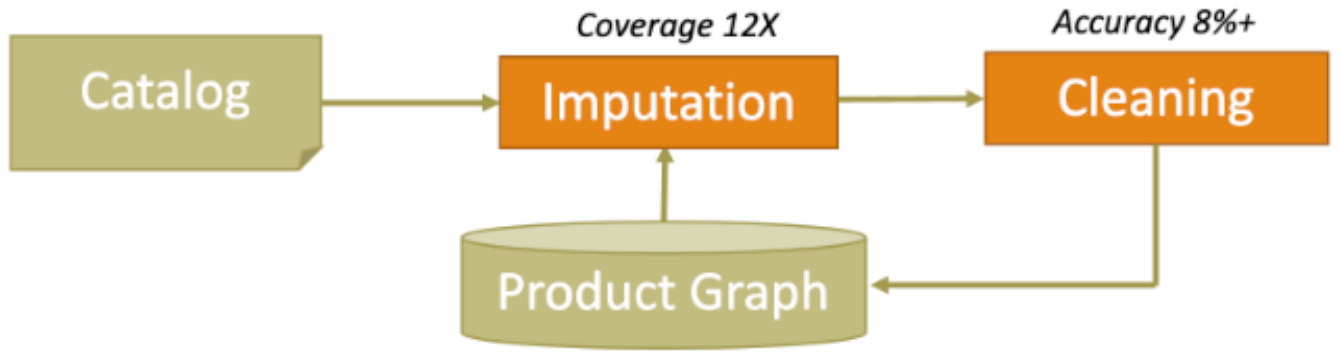}
  
  \vspace{-.85in}
  (b)
  
  \caption{(a) Architecture for constructing an entity-based KG. (b) Architecture for constructing a text-rich KG.}
  \label{fig:architecture}
  \vspace{-.1in}
\end{figure}

To recap, the seed crazy idea behind entity-based KGs is to model the world with entities and relationships, and it faces the challenge that different structured sources express entities and relationships in a heterogeneous way. With knowledge transformation and knowledge integration, major KGs have harvested knowledge from authoritative sources and grown over an order of magnitude over time (\eg, Google KG has grown from 18B triples at launch to over 500B triples\footnote{\url{https://encyclopedia.pub/entry/37713}}). Web extraction from semi-structured websites has also been put in production to supplement long-tail knowledge. Figure~\ref{fig:architecture}(a) depicts key techniques as components in building entity-based KGs. Web-scale knowledge extraction has not been as proliferative as wished, but has inspired research and technical directions to collect long-tail knowledge~\cite{Dong16, LDL+17}.

\eat{
\begin{equation}
  \lim_{n\rightarrow \infty}x=0
\end{equation}

\begin{displaymath}
  \sum_{i=0}^{\infty} x + 1
\end{displaymath}

\begin{equation}
  \sum_{i=0}^{\infty}x_i=\int_{0}^{\pi+2} f
\end{equation}
}

\section{Text-Rich Knowledge Graphs}
\label{sec:textKG}
In many domains like {\em Products, Bioinformatics, Health, Law, Events,} we cannot cleanly model the domain by entities and relationships. We use the {\em Product} domain as an example to illustrate. First, there can be millions of product types, and many of them are overlapping (\eg, {\em fashion swimwear} vs. {\em two-piece swimwear}); thus, defining a clean taxonomy hierarchy is challenging. Second, product attributes are fuzzy and overlapping in nature (\eg, {\em mocha} vs. {\em cappuccino} as flavors, where there could be subtle differences but are also often considered as very similar by most customers); thus, entities may not be the best way to capture them. Finally, products are not strictly {\em named} entities: unlike a person or movie name, product names (\eg, {\em "Onus 2 Colors Highlighter Stick, Shimmer Cream Powder Waterproof Light Face Cosmetics, creamy Self Sharpening Crayon Stick Highlighter"}) are long, verbose, and concatenation of product type and attributes). 

Text-rich KGs are used to model such domains. Instead of setting up clean and strict semantic boundaries between types, relationships, and entities, the {\em majority} of the nodes in text-rich KGs can be just non-canonical {\em texts}. Note that different from entity-based KGs, which often also contain text attributes, here text attributes can be dominant, and it is nearly impossible to extract clean entities from these texts. As such, text-rich KGs are more like bipartite graphs rather than regular connected graphs, with topic entities in the domain on one side of the graph, attribute values (or entities) on the other sides of the graph, connected by attributes (see Figure~\ref{fig:kg-example}(b) as an example).

In the rest of this section, we continue with the {\em Product} domain as an example to describe the techniques, as the aforementioned challenges are best highlighted in the product domain, and e-business has been prevalent in people's lives. Similar techniques have been applied in other text-rich domains~\cite{WAK+21}.

\subsection{{\em Feasibility}: The extraction model}
\label{subsec:prodExtraction}
With the huge semantic ambiguities, it is not hard to imagine that in the product domain, structured data are sparse and error prone~\cite{WAK+21}. The {\em seed crazy idea} behind text-rich KGs is thus to mine structure and model ambiguity from the structure-sparse source data.

Structure mining relies on knowledge extraction, which requires different techniques from those described in Section~\ref{sec:entityKG}, since entities are non-named and attributes can be mostly free texts. We resort to product profiles including product names, descriptions, and bullets, and train {\em Named Entity Recognition (NER)} models to detect patterns that express a particular attribute. Such models, like {\sf OpenTag}~\cite{opentag}, serve as the basis for product knowledge collection.

With the extracted types and attributes, we can mine their relationships (hypernyms, synonyms, etc.) from customer shopping behaviors, such as search, co-view ("customers who viewed this also viewed"), and co-purchase. For example, if users searching for "tea" often buy "green tea", whereas users searching for "green tea" seldom end up buying other types of teas, it hints that "green tea" is a subtype of tea. GNN models have been employed to mine such relationships for types~\cite{octet} and attribute values~\cite{autoknow}. Such methods are also used to establish the substitutes and complements between products~\cite{HZL+20, XMD+21}.

\begin{figure}
  \centering
  \vspace{-.7in}
  
  \includegraphics[width=\linewidth]{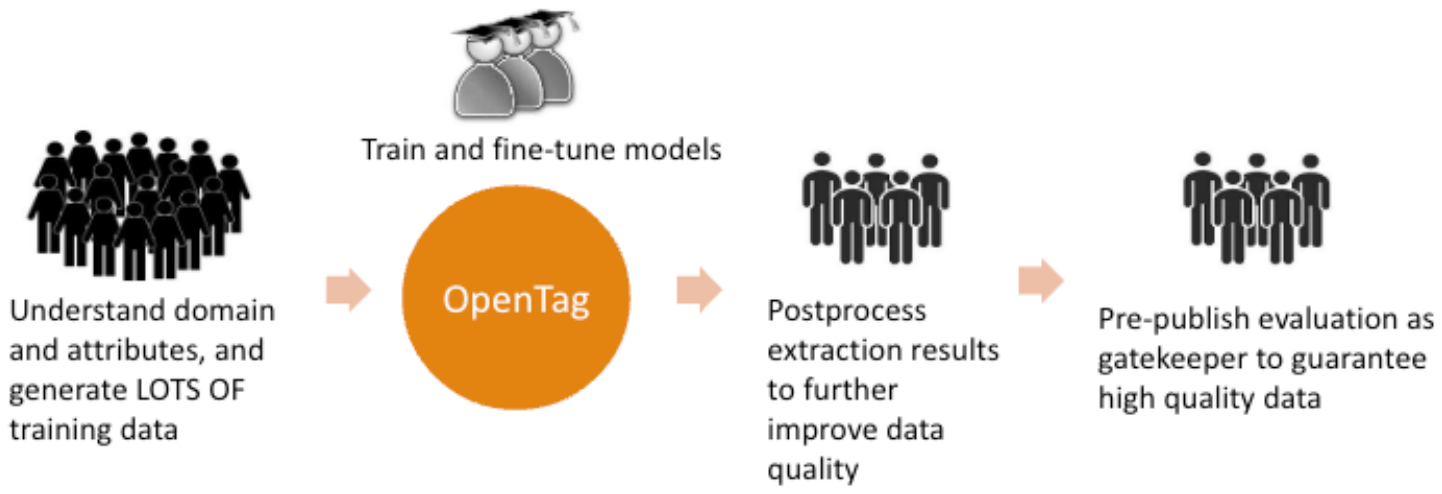}
  
  \vspace{-.6in}
  
  (a)
  
  \vspace{-.4in}
  
  \includegraphics[width=\linewidth]{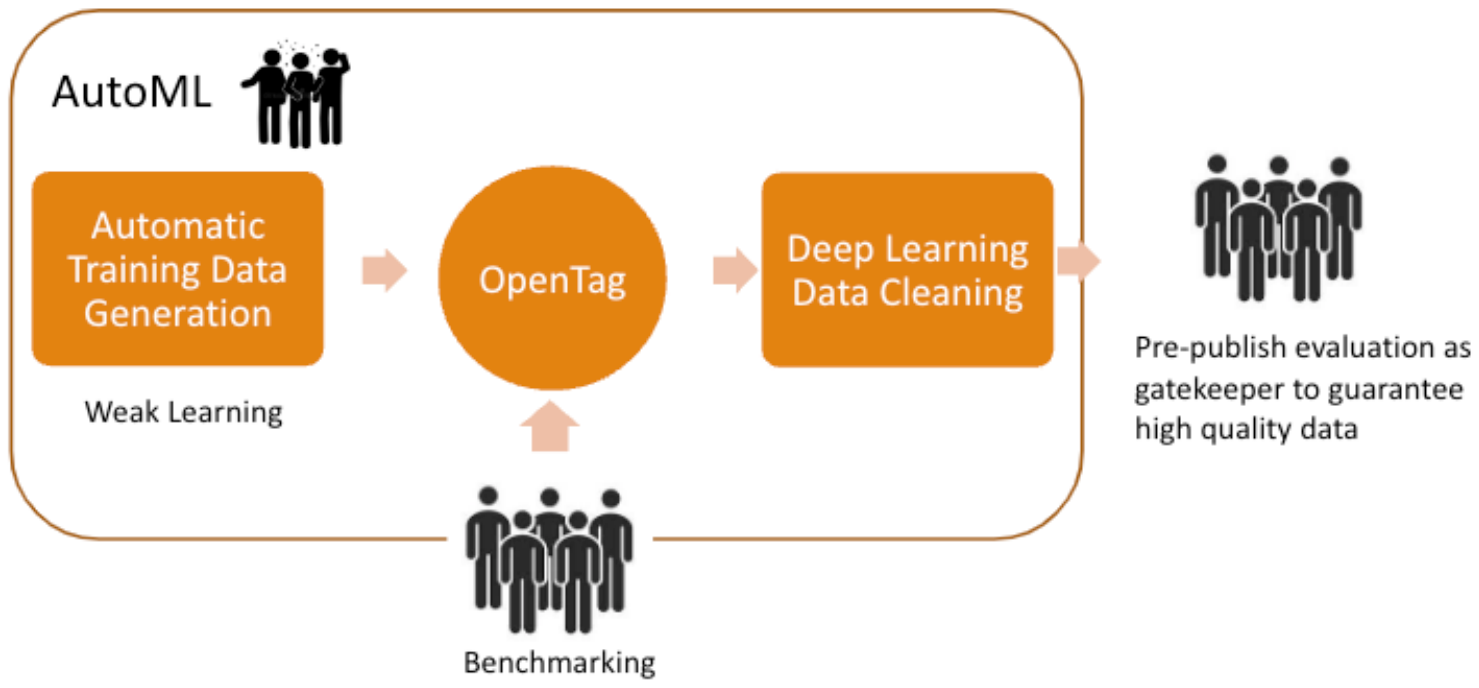}
  
  \vspace{-.6in}
  
  (b)
  
  \caption{(a) Knowledge extraction pipeline to ensure production quality. (b) Knowledge extraction pipeline with reduced manual work.}
  \label{fig:pipeline}
  \vspace{-.2in}
\end{figure}

\subsection{{\em Quality and Repeatability}: The extraction pipeline}
\label{subsec:prodPipeline}
{\bf Quality:} Despite the initial success for NER-based extraction, the quality falls between $85\%-95\%$, so still mediocre. To achieve production quality (\eg, $90\%$), a lot of pre- and post-processing is still needed, as shown in Figure~\ref{fig:pipeline}(a):
\begin{itemize}
    \item Understand the domain and attributes, and generate training data;
    \item Fine-tune hyper-parameters to improve the model;
    \item Postprocess extraction results with rule-based filtering;
    \item Pre-publish evaluation as a gate-keeper to guarantee high quality results on real data in the wild.
\end{itemize}
These methods together allow high-quality extraction, often with accuracy above $95\%$. On the other hand, it introduces a lot of manual work, from labelers, from taxonomists, and from ML engineers and scientists.

\smallskip
\noindent
{\bf Repeatability:} To achieve true repeatability for extractions on different attributes and product types, we need a pipeline that is fairly automatic. The following changes, as described in Figure~\ref{fig:pipeline}(b), aim to remove manual work as much as possible.
\begin{itemize}
    \item Training data are generated by distant supervision, from existing product Catalog. Since Catalog data could be noisy~\cite{autoknow}, we still manually label a small number of instances (tens to hundreds) for benchmarking.
    \item Postprocessing is replaced with deep learning based data cleaning (\eg, transformers, GNNs), leveraging consistency between product descriptions and attributes, between different attribute values of the same product (\eg, snack with sugar in the ingredient is unlikely to be sugar-free), and between products of the same type (\eg, spicy is unlikely to be the flavor of icecreams)~\cite{autoknow, pge, tab-cleaner}. Manual post-processing is only done if ML-based post-processing still cannot achieve the quality bar.
    \item AutoML pipeline is built to reduce model fine tuning efforts and enable non ML-savvies to tune the models.
\end{itemize}
With the above improvements, we observed that the time to train and deploy an extraction model can be reduced from a couple of months to a couple of weeks, allowing steadily generating product knowledge to feed e-business features (information display, product comparison, search, recommendation, etc.). 

\subsection{{\em Scalability}: One-size-fits-all solutions}
\label{subsec:prodScalability}
The product domain can contain millions of product types, thousands of product attributes, and hundreds of languages and locales. Even an efficient pipeline as described in Section~\ref{subsec:prodPipeline} cannot afford to train a model for every combination of product type, attribute, and language. To scale up, we need a solution that is one-size-fits-all.  

The product domain is complex because of the huge type variety; even neighboring product types, such as {\em Coffee} and {\em Tea}, could have quite different attributes, and different vocabularies and patterns for attribute values. One-size-fits-all models need to be able to understand and leverage the subtle differences between types, attributes, and languages when training the models. Once developed, they would significantly increase the volume of product knowledge and thus the business impact. We next give a few examples.

\noindent
{\bf Multi-type extractions:} {\sf TXtract}~\cite{txtract} deepens its understanding of product types for better extraction in two ways. First, it takes the embedding of the product types as part of the input to the model, so the extraction is type-aware. Second, it employs multi-task learning to predict product types in addition to knowledge extraction, for the model to better understand texts related to type semantics. {\sf TXtract} shows that it can train one model for 4K product types, while increasing extraction F-measure by $10\%$ compared to {\sf OpenTag} as a baseline. 

\smallskip
\noindent
{\bf Multi-attribute extractions:} The values for different attributes can be more different than values for the same attribute across different product types, thus requiring slightly different models. {\sf AdaTag}~\cite{adatag}, as an example, takes attribute embeddings as input, and applies Mix of Expert (MoE) and HyperNet to leverage the similarities between the attributes (\eg, {\em flavor} and {\em scent}, though different, share a lot of common vocabularies) in model training. It can train one model for 32 major attributes whereas still improving quality over training one model per attribute. 

\subsection{{\em Ubiquity}: Extraction from broader sources}
\label{subsec:prodMM}
Just as collecting entity-based knowledge, we wish to harvest any knowledge existing for the product domain. The techniques described above focus on text information provided by retailers on one single e-business data source, and we can imagine extending knowledge collection in three directions: product images, customer reviews, and multiple e-business websites. These directions can all lead to a new round of innovations, and we briefly describe early results for multi-modal product knowledge extraction.

Product images (both the visual clues and the texts on products) supplement information not existing in product profiles, or enhance information that is vague or ambiguous in profiles. The {\sf PAM} multi-modal extractor~\cite{pam} employs a multi-modal transformer to attend across texts and images to improve knowledge extraction; in addition, it uses a generative model, adapted according to the product types, to allow extracting values not observed in training data. Experimental results show that it can improve over text extraction by $11\%$ on F-measure.

\subsection{Summary}
To recap, the seed crazy idea behind text-rich KGs is to mine structure and model ambiguity for complex domains, and it faces the challenge that structured data are sparse and noisy. With one-size-fits-all extraction and cleaning (see Figure~\ref{fig:architecture}(b)), Amazon {\sf AutoKnow} system automatically collected 1B knowledge triples over 11K distinct product types, and considerably extended the ontology and improved Catalog quality~\cite{autoknow}. Similar success has been witnessed in other e-business companies~\cite{su-opentag, WYK+20}, and other domains~\cite{WAK+21}.
\section{Dual Neural Knowledge Graphs}
\label{sec:neuralKG}

Comparing with rigorously-structured entity-based KGs, text-rich KGs inject free texts in the structure and thus allow much more flexibility to model complex domains. A natural question to ask next is whether we can completely remove the structure---instead of explicitly modeling knowledge, we capture semantics implicitly such as through embeddings. This question is even more prominent given the recent huge success of LLMs, whose emerging reasoning capabilities~\cite{chain-of-thought} seem to have implied failures of {\em Symbolic AI}~\footnote{One application of symbolic AI is {\em knowledge-based systems}, but here "knowledge" refers to logic rules, instead of structured information as discussed in this paper.}. {\em Will KGs be replaced with LLMs}?

\smallskip
\noindent
{\bf The current:} At the current moment, LLMs clearly have not replaced knowledge graphs. First, training an LLM is expensive. As such, it is hard for LLMs to quickly absorb recent knowledge. For example, GPT-4, released in March 2023, is trained with knowledge up to September 2021, with a 1.5-year lag~\cite{gpt4}. Second, as broadly known, one major problem for LLMs is hallucination of non-existing facts; our recent study~\cite{head-to-tail} shows that for questions that can be answered using DBPedia data, ChatGPT~\cite{chatgpt} has a hallucination rate of $\sim$20\%, and cannot answer $\sim$50\% of them. Finally, LLMs can only learn knowledge when it appears often in the training data; as the same study shows, the accuracy in answering questions involving long-tail facts (questions regarding entities in the bottom 33\% popularity) drops from $\sim$50\% to $\sim$15\%. 

\smallskip
\noindent
{\bf The future:} With the above analysis, we envision a KG that encodes knowledge both in the form of knowledge triples and in the form of LLM embeddings, where the former are easier to use for human understanding and explainability, whereas the latter are easier for machine comprehension. We next elaborate with three subsets of knowledge. 
\begin{itemize}
    \item {\bf Taxonomy:} Taxonomy, or the type hierarchies, is what LLMs are good at capturing. With LLMs, it may not be worth explicitly modeling type relationships (\eg, hypernyms, synonyms, etc.), not to mention manually constructing a very deep and complex hierarchy. So tail taxonomy may best reside at the LLM side.
    \item {\bf Head knowledge:} Training data should be abundant for head knowledge (popular entities and popular attributes) so intuitively there could be a way to teach LLMs head knowledge so they can efficiently address such information needs; in other words, ideally head knowledge reside in both forms. Surprisingly, LLMs can still have a high hallucination rate for head entities (the previously mentioned study shows a hallucination rate of 21\% for DBPedia entities with top-$33\%$ popularity). One important research problem is how to infuse head knowledge into LLMs to enable precise answers to relevant questions, through model training, or through model fine tuning. Early work in this line includes knowledge infusion~\cite{kg-bart, k-adapter}.
    \item {\bf Torso-to-tail and recent knowledge:} With the current techniques for LLM training, LLMs are unlikely to be able to effectively incorporate such knowledge, which is lacking in training data. Thus, such knowledge may best reside as triples. Best serving such knowledge requires knowledge-enhanced LLM, which can effectively decide if such knowledge is required for the conversation, seamlessly plug-in external knowledge sources, and do so efficiently. Early work in this line includes knowledge-augmented LLM~\cite{webgpt, replug, BMH+22}. 
\end{itemize}

How to blend the two forms of knowledge elegantly and how we best address our knowledge needs by leveraging the latest advancements of LLMs remain an open question, and a hot research topic. In addition, how to effectively capture personal knowledge and multi-modal knowledge, leverage them in LLMs to support QA and conversations are even broader research areas. 

\eat{
    \item {\bf Multi-modal knowledge:} We have not mentioned much about knowledge in forms like images, videos, audios, etc.  Finally, we point out the extra challenges presented by multi-modal knowledge, due to the variety of instances within the same category, and due to the variety of appearances of the same entity from different perspectives. 
    \item {\bf Personal knowledge:} Personal knowledge is a kind of tail knowledge, only existing in the person's (digital) memory~\cite{} and thus would not be used to train LLMs. How to leverage personal knowledge in LLMs to support QA on the person's past and improve recommendation for future activities remains a green field. It presents much more challenges: personal knowledge may not be put into structured form to be readily used; the large amount of personal knowledge can cause efficiency issue; and the use of personal memory needs to be privacy-preserving.
}

\section{Reflections: Factors to Industry Success}
\label{sec:reflection}

Before we conclude this paper, we reflect on what are key factors to land {\em crazy science ideas} in industry and lead to real {\em business impact}. As observed from a broad range of research directions, there are two necessary conditions. First, the technique has achieved production quality, or, it is {\em ready}; the bar can be different for different techniques, but high for knowledge correctness, normally 90\% to 99\%. Second, the technique enables significant scale-ups of productivity, or, it is {\em essential}. We now illustrate using KG-relevant topics.

\smallskip
\noindent
{\bf Industry successes:} A few areas have witnessed prosperity in industry, including (1) knowledge-based QA, which improves the way we address people's information needs; (2) entity linkage, which is critical in knowledge integration, as discussed in Section~\ref{subsec:kIntegration}; (3) closed information extraction (ClosedIE), which is critical for scaling up knowledge collection for both entity-based and text-rich KGs, and (4) knowledge cleaning, which is important to filter imprecise knowledge from sources and from extractions. All of these fields satisfy the two conditions: reaching production quality, and increasing productivity significantly to provide better user experiences.

\smallskip
\noindent
{\bf Not-yet successful:} On the other hand, there are research areas that have not seen prevalent industry applications. 

\begin{enumerate}
\item Automatic schema alignment: Schema alignment for a few sources is typically done manually by professional taxonomists to ensure 100\% correctness (as discussed in Section~\ref{subsec:kTransformation}), whereas schema alignment at the web scale is done through ClosedIE. 
\item Knowledge fusion: Integrating knowledge from a few authoritative sources does not encounter too many conflicts, and integrating knowledge from a very large number of sources is not popularly deployed in industry, so the need for fusion is still limited. 
\item Link prediction (aka, knowledge inference): Link prediction has not achieved the quality to reliably add inferred knowledge into KGs; another use of it, to detect incorrect information, has been incorporated into knowledge cleaning techniques. 
\item OpenIE: Extracting knowledge where entities and relations are all texts without any restriction has been considered as a promising way to significantly increase the volume of KGs, but the quality has not been satisfactory for production; recent LLMs have allowed much better ways to capture such {\em fully open} knowledge. 
\end{enumerate}

These techniques miss one of the two factors thus have not seen broad production uses; however, we find that they have inspired new research topics, and sometimes those newly developed techniques "replaced" techniques of the original form. 
\section{Conclusions}
This paper describes generations of knowledge graphs: entity-based KGs, text-rich KGs, and dual neural KGs. In addition, it uses the evolution of KG construction techniques to illustrate the cycle from innovation to production and further to next round of innovation, containing five stages: feasibility, quality, repeatability, scalability, and ubiquity. The recent big success of LLMs shows new directions for knowledge collection and knowledge encoding, which surely will bring knowledge encoding, collection, and mining to the next era and further push business impact to the next level.

\balance 

\begin{acks}
 Sincere thanks to Alon Halevy, Divesh Srivastava, Gerhard Weikum, and Yifan Ethan Xu for the careful reading of the paper and many helpful suggestions to improve the paper.
\end{acks}


\bibliographystyle{ACM-Reference-Format}
\bibliography{base}

\end{document}